\documentclass{aa}  
\usepackage{graphicx}
\usepackage{txfonts}
\usepackage{subcaption}
\usepackage[colorlinks=true, linkcolor=blue, citecolor=blue, urlcolor=blue]{hyperref}
\usepackage{xcolor}
\usepackage{natbib}

\begin{document} 

\titlerunning{High-precision polarization measurements with LEKIDs}
\authorrunning{Savorgnano et al.}

   \title{High-precision polarization measurements with Lumped Element Kinetic Inductance Detectors}

   \author{S. Savorgnano
          \inst{1}\fnmsep\inst{2}\fnmsep\thanks{corresponding author: \nolinkurl{sofia.savorgnano@lpsc.in2p3.fr}}, A. Catalano\inst{1}, J.F. Macías-Pérez\inst{1}, J. Bounmy\inst{1}, O. Bourrion\inst{1}, M. Calvo\inst{2}, O. Choulet\inst{1}, G. Garde\inst{2}, A. Gerardin\inst{2}, M. Kusulja\inst{1}, A. Monfardini\inst{2}, N. Ponthieu\inst{3}, D. Tourres\inst{1} and F. Vezzu\inst{1}}

   \institute{LPSC - IN2P3 - CNRS, 53 Av. des Martyrs, 3800 Grenoble (FR)
         \and Institut Néel - INP - CNRS, 13 Av. des Martyrs, 38000 Grenoble (FR)
        \and IPAG - INSU - CNRS, 414 Rue de la Piscine, 38400 Saint-Martin-d'Hères (FR)}

   \date{Received February 28, 2025; Accepted June 28, 2025}
 
\abstract  
{}
{This work aims to demonstrate that two arrays of Lumped Element Kinetic Inductance Detectors (LEKIDs), when employed in filled array configuration and separated by an external linear polarizer oriented at 45º, can achieve the precision required by next-generation cosmological experiments. The focus here is on validating their ability to meet stringent uncertainty requirements, in particular for polarization angle reconstruction.}
{To achieve this, the uncertainties in the reconstruction of the polarization angle have been characterized in the laboratory using a dedicated closed-circuit 100~mK dilution cryostat. This is optically coupled to a Martin-Puplett interferometer and a custom-designed sky simulator equipped with both photometric and polarized sources, allowing one to reproduce realistic ground-based observation conditions. 
This experimental setup allows us to generate intensity and polarization maps with diffraction-limited resolution, allowing us to determine the polarization angles and their associated uncertainties.}
{The results show performance in line with expectations for the next generation CMB experiments. The polarization angle was reconstructed with an uncertainty of 6.5~arcmin.}

   {}


   \maketitle

\section{Introduction}

The Cosmic Microwave Background (CMB) provides an extraordinary window into the Early Universe, offering profound insights into its structure, evolution and fundamental parameters (\citealt{hudod}). Over the past few decades, high-precision CMB measurements have revolutionized our understanding of cosmology, confirming the Big Bang model and supporting the inflationary paradigm (\citealt{planck2020a}, \citealt{planck2020b}, \citealt{act}, \citealt{spt}). CMB polarization, in particular, encodes valuable information about Early Universe physics, including energy scales, particle interactions, and the nature of primordial fluctuations that seeded cosmic structures observed today (\citealt{rees}; \citealt{kovac}). The polarization signal can be decomposed into two distinct polarization patterns: E-modes, which result from scalar density perturbations and exhibit symmetric, radial patterns, and B-modes, characterized by curl-like patterns generated by tensor perturbations such as inflationary gravitational waves (\citealt{huwhite}).
\\
Current and upcoming experiments, such as ground-based observatories like Simons Observatory (SO) (\citealt{ade}) and CMB-S4 (\citealt{cmbs4}) or satellites like LiteBIRD (\citealt{litebird}) and PICO (\citealt{hanani}), aim to push the boundaries of sensitivity in detecting CMB polarization. 

The detection of inflationary gravitational waves encoded in B-modes would confirm the inflationary model and provide a pathway to explore physics beyond the Standard Model (\citealt{guth}; \citealt{polnarev}, \citealt{kam}). The energy scale of inflation is parameterized by the tensor-to-scalar ratio $r$, with current upper limits from the combined Planck and BICEP/Keck data constraining $r < 0.036$ (\citealt{tristram}). Theoretical models predict a wide range of $r$ values: large-field inflation models can yield $r > 10^{-2}$, while models such as the Starobinsky model predict $r$ values closer to $10^{-3}$. Many other models predict $r$ values as small as $10^{-5}$ or lower. Achieving the sensitivity required to probe this wide range of $r$ values necessitates significant advancements in both instrumentation and data analysis techniques (\citealt{verges}).
\\
Experimental biases can introduce systematic errors, making it challenging to separate cosmological signals from Galactic foregrounds and leading to leakage between E-modes and B-modes. Specifically, it has been estimated (\citealt{rosset}) that an uncertainty in the determination of the polarization angle $\geq 0.1^\circ$ can cause significant leakage between E-modes and B-modes, particularly in the multipole range $80<\ell<100$, assuming a targeted sensitivity of $r \sim 10^{-3}$. This assessment is specific to experiments targeting small angular scales and moderate constraints on the B-mode signal, and may not directly apply to experiments covering large angular scales or imposing more stringent constraints on $r$. Therefore, one critical challenge lies in the precise calibration of the polarization angles. Self-calibration methods have been employed to address such issues (\citealt{keating}), but these rely on assumptions about the CMB polarization signal, limiting the potential for new discoveries such as cosmic birefringence (\citealt{minami}; \citealt{diego2}; \citealt{jost}) or polarization rotation caused by parity-violating physics (\citealt{carroll}; \citealt{pospelov}) or yet primordial magnetic fields (\citealt{carroll}; \citealt{diego}).
The complexity of modeling Galactic polarization power spectra, influenced by the alignment of dust grains with the magnetic field orientation, further complicates this calibration process (\citealt{ritacco}; \citealt{vacher}).
\\ 
For these reasons, future experiments demand a polarization angle calibration with a precision at least ten times better than the $\sim$1° achieved by the Planck mission (\citealt{rosset}). Efforts have been initiated to address this need by developing external references for high-precision absolute polarization angle calibration. Notable among these is the COSmological Microwave Observations Calibrator (COSMOCal) project, which has demonstrated solid results in laboratory tests (\citealt{cosmocal}). While COSMOCal could provide the much-needed external reference, meticulous calibration tailored to the specific characteristics of each instrument remains equally critical to achieving the required precision.
\\
In millimeter-wave instrumentation, the focal plane architecture plays a critical role in determining the optical coupling efficiency, field of view and instrument scalability. For example, a horn-coupled focal plane uses feedhorns to guide incoming radiation to the detectors, providing excellent beam control and low crosstalk between KIDs (\citealt{posada}), but comes at the cost of increased complexity, bulk and limitations in packing density, which restricts the number of detectors per unit area. In contrast, a filled-array focal plane directly couples radiation to detectors without intermediate optical elements (\citealt{monfardini}, \citealt{doyle}). This configuration enables higher detector filling factor, making it well-suited for large-format arrays, higher overall quantum efficiency and a simplified fabrication process.
The NIKA2 instrument is installed on the IRAM 30~m telescope and is utilizing filled-array LEKIDs. LEKIDs are highly sensitive superconducting detectors that use the kinetic inductance effect to measure incoming radiation. Unlike other types of KIDs, LEKIDs integrate the resonator and the absorber into a single structure, allowing the same physical element to both absorb incoming radiation and resonate at a specific frequency. This simplifies the fabrication and optimizes their coupling efficiency for millimeter and sub-millimeter wavelengths (\citealt{doyle}). NIKA2 has demonstrated outstanding performance in mm-wavelength polarimetry. It achieves a Noise Equivalent Flux Density (NEFD) of $\sim 33$~mJy$\cdot$s$^{1/2}$ for total intensity and of 20~mJy$\cdot$s$^{1/2}$ in polarization for $Q$ and $U$ at 1~mm (\citealt{comrep}), thanks to the high speed modulation (approximately 3~Hz) provided by the continuously rotating Half Wave Plate (HWP). The instrumental leakage remains below 3\%, and the statistical uncertainty on the polarization angle reconstruction is limited to $\pm 1^\circ$ (\citealt{ritacco17}; \citealt{adam}; \citealt{catalano}). This latter result is of particular importance for this study, as it sets the current baseline for LEKIDs used for polarization, either astrophysical or cosmological. The best results have been obtained during the observations of the Crab Nebula (\citealt{ritacco22}), still representing a reference for any polarization calibration.  
\\
However, achieving the performance required for the next generation cosmological experiments for CMB polarization of $<$ 0.1$^\circ$ remains challenging. A significant limitation arises from environmental conditions and the optical configuration commonly encountered in ground-based facilities, exemplified here by the 30~m IRAM telescope Nasmyth cabin, which was not originally optimized for polarization measurements. While the experimental setup described in this paper does not directly assess the limitations due to the telescope environment, future upgrades to the current setup could enable such an evaluation. Addressing these limitations would enhance the precision of KID-based polarization measurements, offering valuable insights for both cosmology and Galactic foreground studies.
\\
This study aims to demonstrate the performances of a measurement technique for linear polarization using LEKIDs in filled-array configuration. In this setup, two arrays are arranged perpendicular to each other, the incoming beam split on the two by a 45º-oriented linear polarizer. The results presented here were obtained in a band centered at 150~GHz, which is particularly relevant for studies of CMB polarization. These findings reinforce the validity of the measurement approach, establishing a robust foundation for its application in future instruments dedicated to observing CMB polarization and disentangling Galactic foregrounds.
\\ \\
This article is structured as follows: Sect.~\ref{sec:exp} describes the experimental setup and the key elements of the measurement configuration; Sect.~\ref{sec:opt_sim} presents the characterization of the test bench; Sect.~\ref{sec:perf_test} presents the camera's results in terms of the electrical, photometric and polarimetric characterization; Sect.~\ref{sec:cosmo} discusses the impact of the obtained results in the context of CMB B-modes detection; Sect.~\ref{sec:concl} draws the conclusions of this work, underlining the main achievements and lessons learned, alongside the perspectives for further development. 

\section{The experimental setup}
\label{sec:exp}

The experimental setup is composed of two main parts: the test bench and the LEKIDs camera. These two parts are described separately, respectively in Subsect.~\ref{subsec:testbench} and Subsect.~\ref{subsec:lekids_camera}.

\subsection{The test bench}
\label{subsec:testbench}

\begin{figure*}
    \centering
    \includegraphics[width=\textwidth]{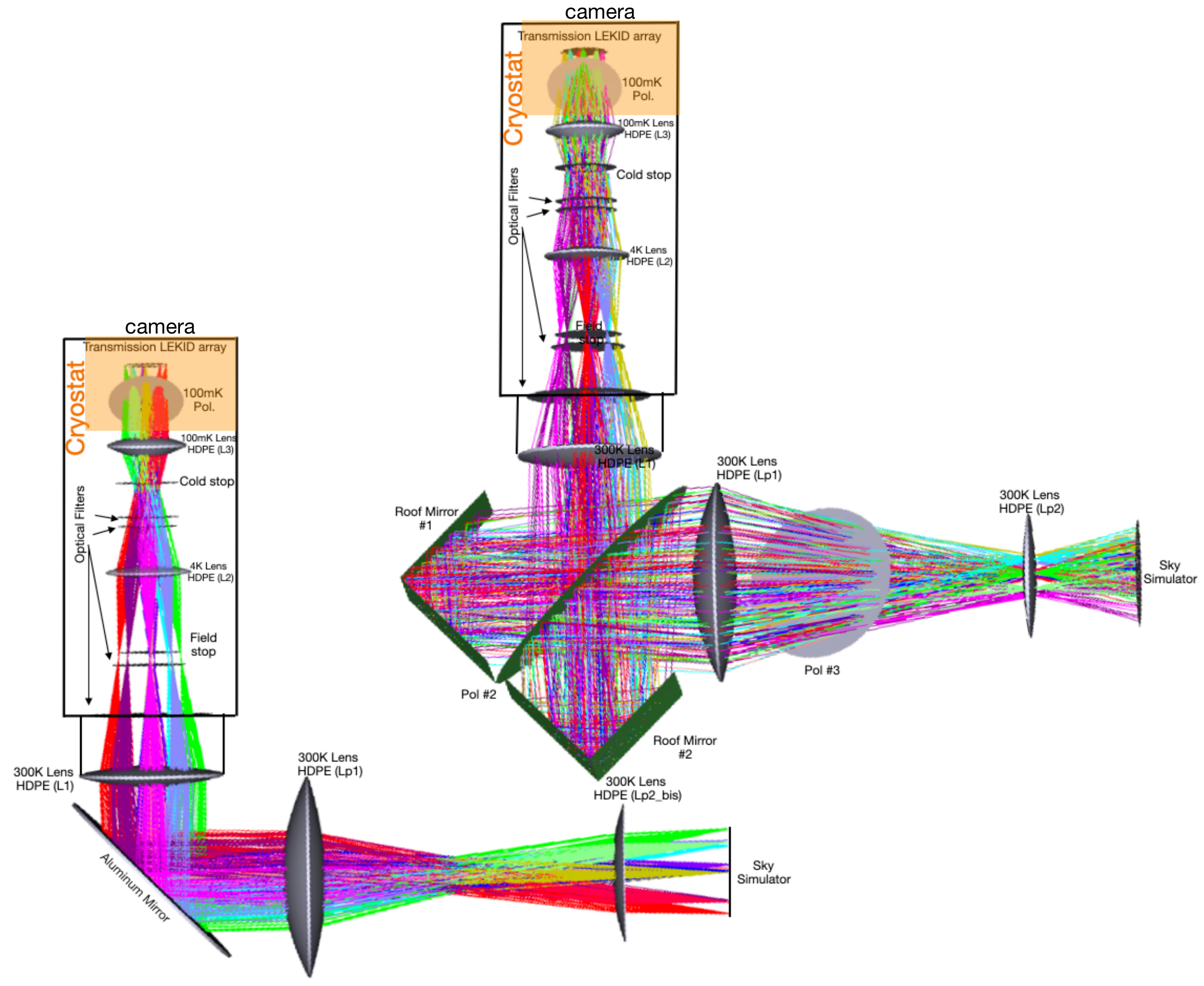}
    \caption{Zemax representation of the experimental setup in the two implemented configurations: left - photometric and polarimetric configuration; right - spectrometric configuration employing the Martin-Puplett interferometer. All the elements composing the setup are individually labeled in the figure. The orange shaded regions highlight the camera within the cryostat.}
    \label{fig:zemax_sys}
\end{figure*}

The test bench is composed of a closed-cycle $^3$He - $^4$He dilution cryostat designed for optical measurements. The cryostat is optically coupled to a sky simulator, which is a tool that allows us to mimic the emission from the atmosphere in on-sky observations. The sky simulator can be coupled to the main cryostat using a 45º-oriented linear polarizer and a Martin-Puplett interferometer (as shown in the right panel of Fig.~\ref{fig:zemax_sys}) for spectroscopic characterization or using a 45º-oriented flat mirror for photometric and polarimetric characterization (as shown in the left panel of Fig.~\ref{fig:zemax_sys}).

The entire system is optimized to operate under a low optical background, typical of a ground-based instrument at 150~GHz. This optical background is achieved by coupling the camera to a sky simulator. The optical background can be adjusted by regulating the cold stop aperture within the camera (\citealt{catalano}).
The sky simulator is capable of performing both total intensity and polarization measurements, thanks to interchangeable optical sources positioned in front of the cold black body. Each subsystem is detailed below. 

\subsubsection{Optical coupling with the sky simulator} 
The coupling between the camera and the sky simulator is achieved using two HDPE lenses and two possible optical configurations, depending on the type of measurement to be performed (see Fig.~\ref{fig:zemax_sys}). For spectral measurements, a Martin-Puplett interferometer is used (\citealt{fasano}, \citealt{macias}), capable of performing interferograms with a path length up to 80~mm, providing a spectral resolution of about 1~GHz. For total intensity measurements, the camera is directly coupled to the sky simulator via a 45$^\circ$ flat mirror. The reduction in optical path length compared to the spectral configuration is compensated by replacing one of the two HDPE lenses.

\subsubsection{Sky simulator}
The sky simulator consists of two primary components: a cold black body and various interchangeable optical sources. The cold black body is created by cooling a large black disk (270~mm in diameter) to approximately 16~K inside a dedicated pulse tube cryostat. The cryostat includes a 50~K screen equipped with an IR filter made of ZITEX G110. The window is made of HDPE and has a diameter of 300~mm. The optical sources are mounted on a Kapton membrane with a diameter of 400~mm and a thickness of 50~$\mu$m. These sources are placed at 300~K, in front of the cold black body. The sources used for photometric and polarization characterization include:

\begin{itemize}
    \item Photometric point source: a high-emissivity disk with a diameter of 3~mm, attached to the Kapton membrane. The optical source produces a beam on the focal plane with FWHM $\sim$14~mm. This means that a 3~mm disk produces a contrast of $\sim$15~K. 
    \item Extended polarized source: a 30~mm diameter polarizer fixed to the Kapton membrane, acting as a fully polarized extended source. It generates a temperature contrast of approximately 150~K, simulating a polarized nebula from an observational perspective.
\end{itemize}

The optical sources can be moved vertically and horizontally within a range of $\pm$150~mm around the optical axis using two DC geared motors and linear tables, which provide a resolution of 0.1~mm. A third motor enables rotational motion at speeds of up to 1~Hz with an accuracy of 0.1$^\circ$. Position control is managed via controllers and encoders, ensuring precise alignment. A GPS clock operating at 10~MHz ensures precise synchronization between the data streams of the KIDs and the pointing system. Additionally, a pulse-per-second (PPS) signal establishes a unified starting point for data acquisition.

\begin{figure}
    \centering
    \includegraphics[width=\hsize]{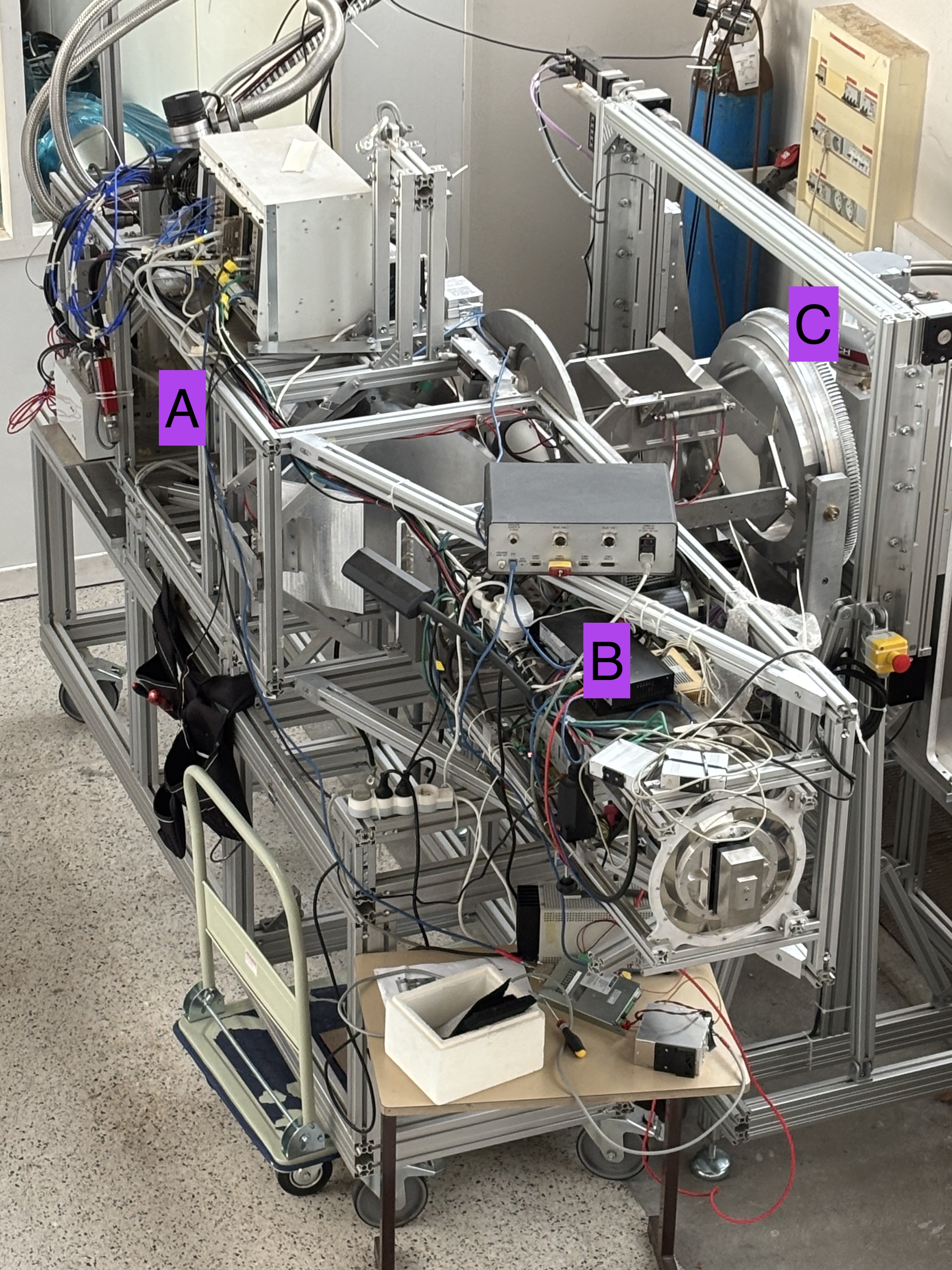}
    \caption{Picture of the experimental setup in the cryogenic lab of the Laboratoire de Physique Subatomique et Cosmologie (LPSC) of Grenoble: A marks the 100~mK cryostat, B marks the Martin-Puplett interferometer and C marks the sky simulator.}
    \label{fig:sky_sim}
\end{figure}

The fully assembled experimental setup is illustrated in Fig.~\ref{fig:sky_sim}, while Tab.~\ref{tab:setup_char} summarizes the main characteristics and requirements for this study.

\subsection{The LEKIDs receivers}
\label{subsec:lekids_camera}

On the other end of the experimental setup with respect to the sky simulator, the actual receivers are installed. The receivers are a pair of LEKIDs arrays, arranged in a specific configuration. The camera itself, the arrays' configuration and the coupling with a Vector Network Analyzer (VNA) are detailed in the following subsections.

\subsubsection{The camera}
The two LEKIDs arrays are mounted in the coldest stage of the cryostat, which maintains a base temperature of approximately 130~mK with 0.1~mK stability. The cryostat is designed to support two independent RF channels, each equipped with a cryogenic low-noise amplifier. Since LEKIDs are sensitive to magnetic fields, three magnetic shields have been implemented to mitigate this noise source: a $\mu$-metal enclosure at 300~K, a superconducting lead screen at the 1~K stage, and an aluminum screen at the 100~mK stage.

Within the camera, a cold stop is created at the 100~mK stage, while a field stop is positioned at the 4~K stage. These optical components are assured by three High-Density Poly Ethylene (HDPE) lenses, located at the 100~mK (L3), 4~K (L2) and 300~K (L1) cryogenic stages. 

To limit the radiative load from the optical access, a sequence of filters are placed at various cryogenic stages. These include IR-blocking thermal filters at the warmest stages, as well as metallic multi-mesh low-pass filters at lower temperatures (\citealt{catalano2015}, \citealt{monfardini2015}). A blackened baffle is installed at the 4~K stage, immediately after the field stop, to minimize stray-light contamination.

\subsubsection{LEKIDs array configuration} 

\begin{figure}
    \centering
    \includegraphics[width=\hsize]{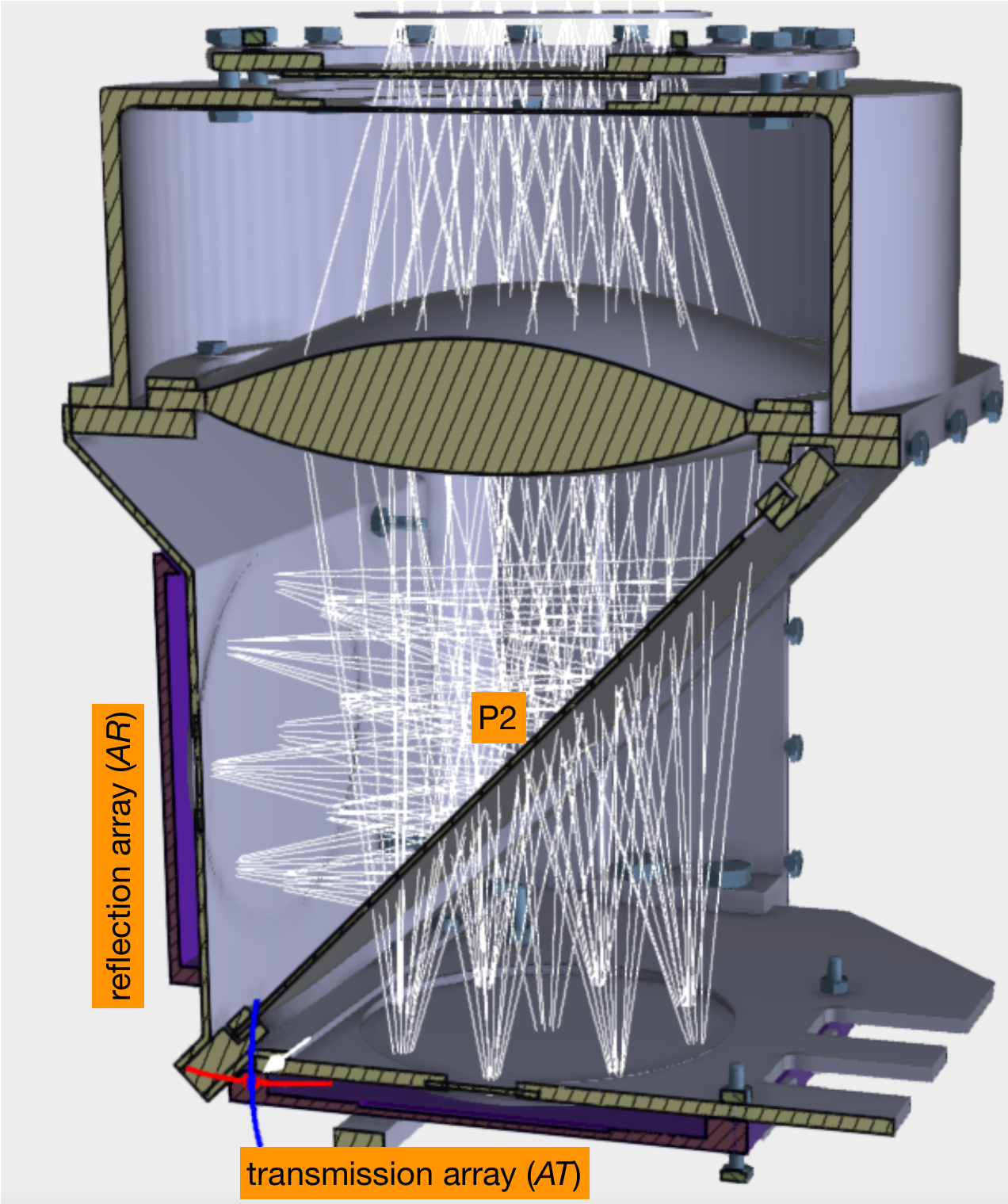}
    \caption{CAD design of the 100~mK stage, where the linear polarizer (P2) and the two perpendicular arrays are located.}
    \label{fig:zemax_focplane}
\end{figure}

The focal plane comprises two single-polarization sensitive LEKIDs arrays positioned perpendicularly to each other, named $AT$ for $transmission~array$ and $AR$ for $reflection~array$. Each array is designed to contain 418 pixels, but due to the typical sub-band layout of the electronic readout system, only 317 pixels per array are actually read, reducing the number of available pixels below the design specification. A linear polarizer (P2), oriented at 45$^\circ$, is placed to split the incoming beam between the two arrays (see Fig.~\ref{fig:zemax_focplane}). The action of this polarizer is effectively to introduce a phase difference of 90$^\circ$ between the two detected polarization components. 

The arrays used in this study were fabricated in the microfabrication facility PTA (Plateforme Technologique Amont\footnote{\url{http://pta-grenoble.com/fr/}}). Each individual pixel is a kinetic inductance detector and all the pixels are connected to a quarter-wave coplanar waveguide (CPW). One end of the resonator is coupled to the measurement transmission line, while the other end is grounded. The fabrication process has been thoroughly described in several previous studies, including \citealt{monfardini2009} and \citealt{goupy}. 

The pixels of both arrays are back-illuminated through the silicon wafer. To maximize photon absorption, we employ a back-short cavity located at an optimized distance, which is calculated individually for each array to account for the wafer thickness variations.

To set the high-frequency cut-off, metal-mesh based Low Pass Edge filters are used, while the low-frequency cutoff is achieved through the superconducting cutoff of the thin aluminum film. The polarizer is a wire grid made of copper wires with a pitch of 50~$\mu$m, realized on a 30~$\mu$m thick Kapton membrane.

\subsubsection{Coupling with the VNA}

The VNA is a key component of the experimental setup, playing a crucial role in electrical testing. While we will not delve into the specific details of the VNA itself, we will describe how the camera is coupled to it. Each array is connected to the VNA one at a time via coaxial cables, which run directly from the coldest stage of the cryostat, where the arrays are housed, to the input and output channels of the VNA. This ensures a reliable electrical connection for precise testing. The optical configuration remains the same as that used for photometric characterization; however, in this case, the readout system exclusively relies on the VNA for measurements. 

\begin{table*}[ht]
\caption{Main characteristics and requirements of the experimental setup, divided into test bench and camera specifications.}
\centering
\begin{tabular}{l p{0.7\textwidth}}
\hline
\hline
\multicolumn{2}{c}{Test bench characteristics} \\ 
\hline
Component                           & Description \\ 
\hline
Sky simulator's black body          & 270~mm diameter black disk cooled to 16~K \\ 
Cryostat                            & Custom pulse tube cryostat cooled to 130~mK \\ 
Window                              & HDPE, 300~mm diameter \\ 
Optical sources                     & Mounted on 50~$\mu$m Kapton membrane (400~mm diameter): 
                                      \begin{itemize}
                                          \item High-emissivity disk (3~mm diameter) for photometric point source
                                          \item 30~mm polarizer for extended polarized source
                                      \end{itemize} \\ 
Source movement range               & $\pm$150~mm horizontally and vertically \\ 
Motor control                       & DC geared motors with 0.1~mm resolution and rotation up to 1~Hz \\ 
Position synchronization            & Rubidium 10~MHz clock \\ 
Lenses                              & Two HDPE lenses for coupling camera and sky simulator \\ 
Stray-light reduction               & Blackened baffle at 4~K stage \\ 
\hline
\multicolumn{2}{c}{Camera characteristics} \\ 
\hline
Component                           & Description \\ 
\hline
Temperature stability               & 0.1~mK  \\ 
Arrays                              & CPW LEKIDs \\ 
Illumination                        & back illuminated \\
Configuration                       & filled-array \\
\hline
\end{tabular}
\label{tab:setup_char}
\end{table*}

\section{Test bench characterization}
\label{sec:opt_sim}

The test bench characterization focuses on the precise estimation of the effective temperature of the sky simulator and the associated optical load on the detectors. These parameters are critical for assessing the sensitivity of the setup, optimizing the detector's performance, and ensuring that the noise contributions are well understood and minimized.

\subsection{Effective sky simulator's temperature estimation}
\label{sec:sky_sim_temp}

The temperature of the sky simulator is a critical parameter in the characterization of our test bench. Many of the models and measurements presented rely on an accurate estimation of this temperature to derive meaningful physical results. The sky simulator’s temperature is monitored using a carbon resistance thermometer installed on the cold Stycast plate, a structure made of a mixture of Stycast epoxy, silicon grains and lampblack. These materials allow the plate to maintain a stable and controlled temperature that simulates the cold background of the sky. The measured value, $\sim$16~K, does not correspond to the effective background temperature perceived by the detectors. This discrepancy arises from the additional emission from 300~K and the sky simulator HDPE windows.  

To estimate the equivalent sky simulator temperature, we analyze three VNA scans performed under different optical loads, using only the transmission array $AT$ for simplicity. The optical loads correspond to:  

\begin{itemize}
    \item $T_1$: the nominal sky simulator temperature ($T_{\text{sky}}$),
    \item $T_2$: the nominal sky simulator temperature plus a variation ($T_{\text{var}}=20~\text{K}$),
    \item $T_3$: 300~K, corresponding to an Eccosorb layer placed in front of the sky simulator.
\end{itemize}

The responsivity of the array, $\text{R}$, is derived from the resonance frequency variation $\Delta f$ between $T_2$ and $T_1$ as:

\begin{equation}
    \text{R} = \frac{\Delta f}{T_{\text{var}}} = \frac{(6\pm0.3)~\text{kHz}}{20~\text{K}} = (0.31\pm0.03)~\text{kHz/K}.
\end{equation}

The effective sky simulator temperature, $T_{\text{sky}}$, is computed using:

\begin{equation}
    T_{\text{sky}} = T_3 - \frac{f_3 - f_1}{\text{R}},
\end{equation}

where $f_3 - f_1 = (83 \pm 1)~\text{kHz}$ is the resonance frequency shift between $T_3$ and $T_1$. Substituting the values:

\begin{equation}
    T_{\text{sky}} = 300~\text{K} - \frac{(83\pm1)~\text{kHz}}{(0.31\pm0.03)~\text{kHz/K}} = (22.3 \pm 0.8)~\text{K}.
\end{equation} 

This value reflects the combined contributions from the sky simulator and the entrance HDPE lens. The uncertainty in the temperature is derived from the statistical distribution of the KID data. Specifically, the resonance frequency shifts, which are used to calculate the temperature, are averaged across the pixels in the array. The associated uncertainty is then determined as the standard deviation of this average. It is important to note that the reference temperatures provided by thermometers are treated as absolute values without associated uncertainties, as it was not possible to reliably quantify their uncertainty.

\subsection{Optical load estimation}

In order to compare the sensitivity of the detectors to their equivalent photon noise, an optical simulation was performed to estimate the incident power on each KID. The optical load on a detector, $W_{\rm pixel}$, is the sum of the optical power generated by the sky simulator and all self-emission contributions, as shown in Eq.~\ref{eq:model_emi}:

\begin{equation}
\begin{aligned}
W_{\text{pixel}} &= W_{\text{sky sim}} + W_{\text{self emission}} \\
&= \int_0^\infty \eta(\nu) A\Omega(\nu) S(\nu) BB(T_{\text{sky sim}}, \nu) \, d\nu \\
&\quad + \int_0^\infty \eta(\nu) A\Omega(\nu) S(\nu) BB(T_{\text{self~emission}}, \nu) \, d\nu\, ,
\end{aligned}
\label{eq:model_emi}
\end{equation}

where:

\begin{itemize}
    \item $\eta(\nu)$ is the overall transmission, mostly defined by the transmission of the 5 HDPE lenses; the transmission of each single lens is calculated as:
    \begin{equation}
    \eta_{\text{lens}}(\nu) = (1 - 2\epsilon_{\text{HDPE}}) e^{-2 \pi \nu B p \frac{n_{\text{HDPE}}}{c}}\, ,
    \end{equation}
    with $\epsilon_{\text{HDPE}}$ and $n_{\text{HDPE}}$ being the dielectric constant and the refraction index of the HDPE respectively, $p$ the thickness of the lens, and $B$ an empirically measured parameter. The overall transmission over the selected bandwidth is $\eta$ = 0.8. 
    
    \item A$\Omega (\nu)$ is the throughput of each pixel and it is a geometric property of the optical system. For a diffraction-limited system, it depends on the effective collecting area A and the solid angle $\Omega$ subtended by the diffraction-limited beam, and it can also be expresses as:

    \begin{equation}
    A\Omega(\nu) = \frac{\pi}{4} (F\lambda)^2=0.54 ~~~\text{with} \quad F\lambda = \frac{Fc}{\nu}=0.83\, ,
    \end{equation}

    where $F$ is the focal ratio (also referred to as the F-number), meaning the ratio of the focal length to the aperture diameter of the system.

    \item $D(\nu)$ is the bandpass normalized to one (Fig.~\ref{Fig:band_2mm})

    \item $BB(T, \nu)$ is the black body brightness as a function of the working temperature, $T$, and frequency, $\nu$
\end{itemize}

By performing this calculation using the actual parameters of our setup, we find that at the working temperature of the sky simulator of $22.3~\text{K}$ (derived in Sect.~\ref{sec:sky_sim_temp}), the total incident power per pixel is $(20 \pm 5)~\text{pW}$. The 30\% uncertainty reflects the absorption factor of the optical components, which is based on experimental estimates due to the lack of precise measurement tools.

From this estimate, we can compute the photon Noise Equivalent Power ($\rm{NEP}_{\rm phot}$) at the central frequency of the bandpass (150~GHz) as:

\begin{equation}
    \mathrm{NEP_{phot}} = \sqrt{2 h \nu_{\text{central}} W_\mathrm{pixel}} = (6.6\pm1.5) \times 10^{-17} \, \mathrm{W/\sqrt{Hz}}\, .
\end{equation}

The photon NEP can be directly compared to the total NEP, which is composed of multiple noise sources. In order to refer to photon-noise-dominated detectors, we want the following expression to be satisfied:

\begin{equation}
    \mathrm{NEP_{goal}} \lesssim \sqrt{2} \, \mathrm{NEP_{phot}}\, .
    \label{Eq:NEP_phot}
\end{equation}

In this way, the biggest contribution to the system noise is the photon noise itself.

\section{Camera's performances}
\label{sec:perf_test}

We conducted extensive measurements to evaluate both the total intensity and polarization performances of the LEKIDs arrays and of our test bench in general.

The primary goals of these measurements were:

\begin{enumerate}
    \item Electrical characterization: identification of resonances and estimation of the quality factors using VNA frequency sweeps (Subsect.~\ref{sec:elec}).
    \item Photometric and spectral characterization: estimation of the working bandpass using the Martin-Puplett interferometer, assessment of noise properties and optical responsivity, and characterization of the photometric response to a point source, including beam properties (Subsect.~\ref{sec:photo}).
    \item Polarimetric characterization: determination of the polarization angle along with its associated uncertainty (Subsect.~\ref{sec:polar}).
\end{enumerate}

This workflow, summarized in Tab.~\ref{tab:char_mode}, was designed to demonstrate that our test bench can achieve the accuracy required by current cosmological experiments. The primary objective of the polarized measurements is to ensure that the error associated with the estimated polarization angle is $\leq$ 0.1$^\circ$. The photometric and polarimetric analyses were performed using 30\% of the best-performing pixels, selected based on their low noise levels, expected beam shapes and high responsivity. The rationale for the choice of this subset of pixels lies in the focus of the study: to validate an experimental method rather than to assess the uniformity of results across the entire array. Achieving the required performance with this selected group of KIDs is sufficient to demonstrate the method’s feasibility.

\begin{table}
\caption{Characteristics of the camera and its coupling configurations.}
\centering 
\begin{tabular}{lcl} 
\hline\hline 
Characteristic & Mode & Configuration \\ 
\hline 
Nº of resonances &  mirror & VNA \\ 
Quality factors &  mirror & VNA \\
Bandpass & polarizer & Martin-Puplett \\ 
Noise &  mirror & photom. source \\
Optical responsivity &  mirror & photom. source/VNA \\ 
Beam properties &  mirror & photom. source \\
Polarization angle &  mirror & ext. polar. source \\
\hline 
\end{tabular}
\label{tab:char_mode}
\end{table}

\subsection{Electrical characterization}
\label{sec:elec}

\begin{figure}[ht]
    \centering
    \begin{subfigure}[b]{\linewidth}
        \centering
        \includegraphics[width=\linewidth]{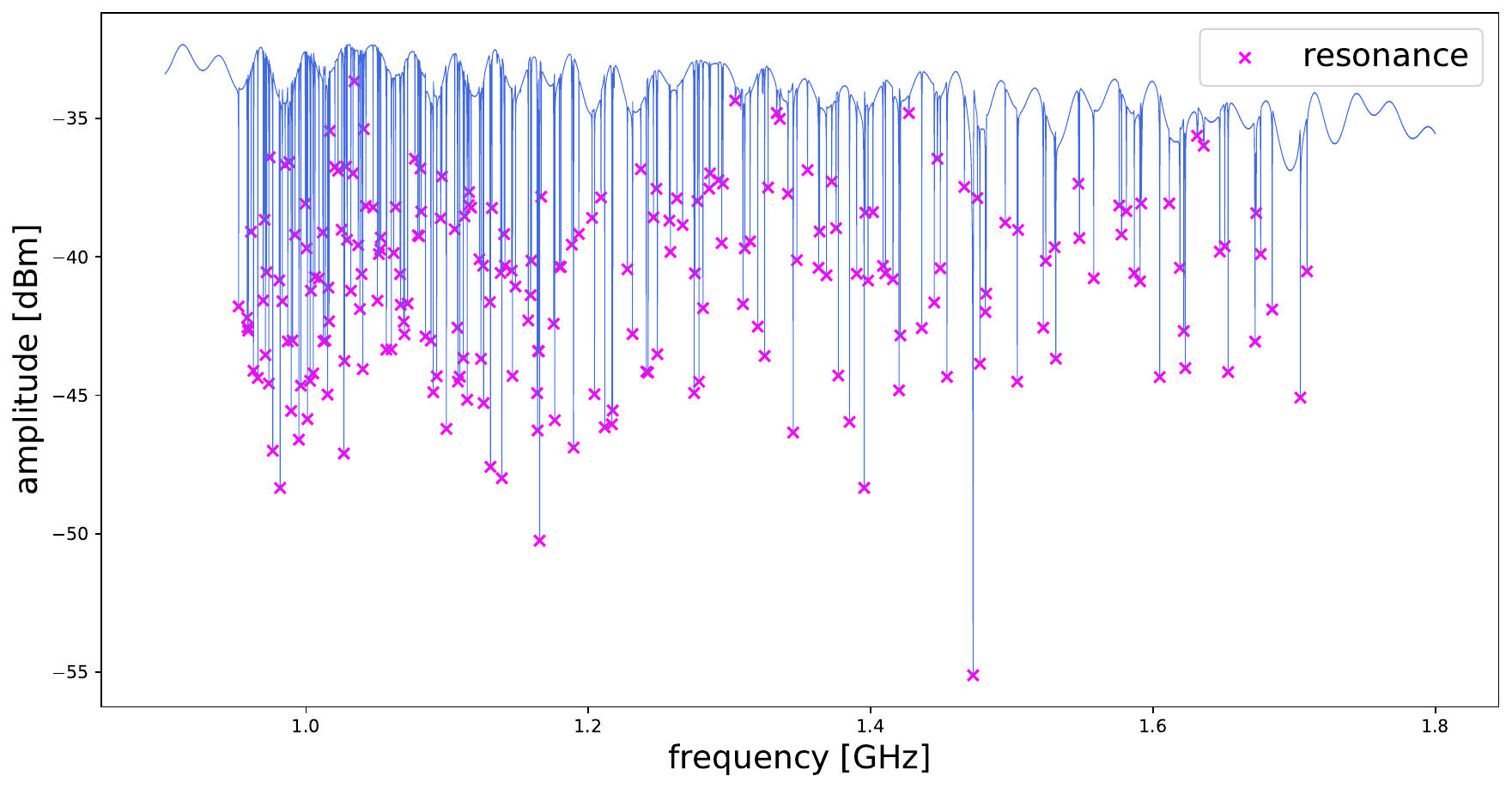}
    \end{subfigure}
    \vfill
    \begin{subfigure}[b]{\linewidth}
        \centering
        \includegraphics[width=\linewidth]{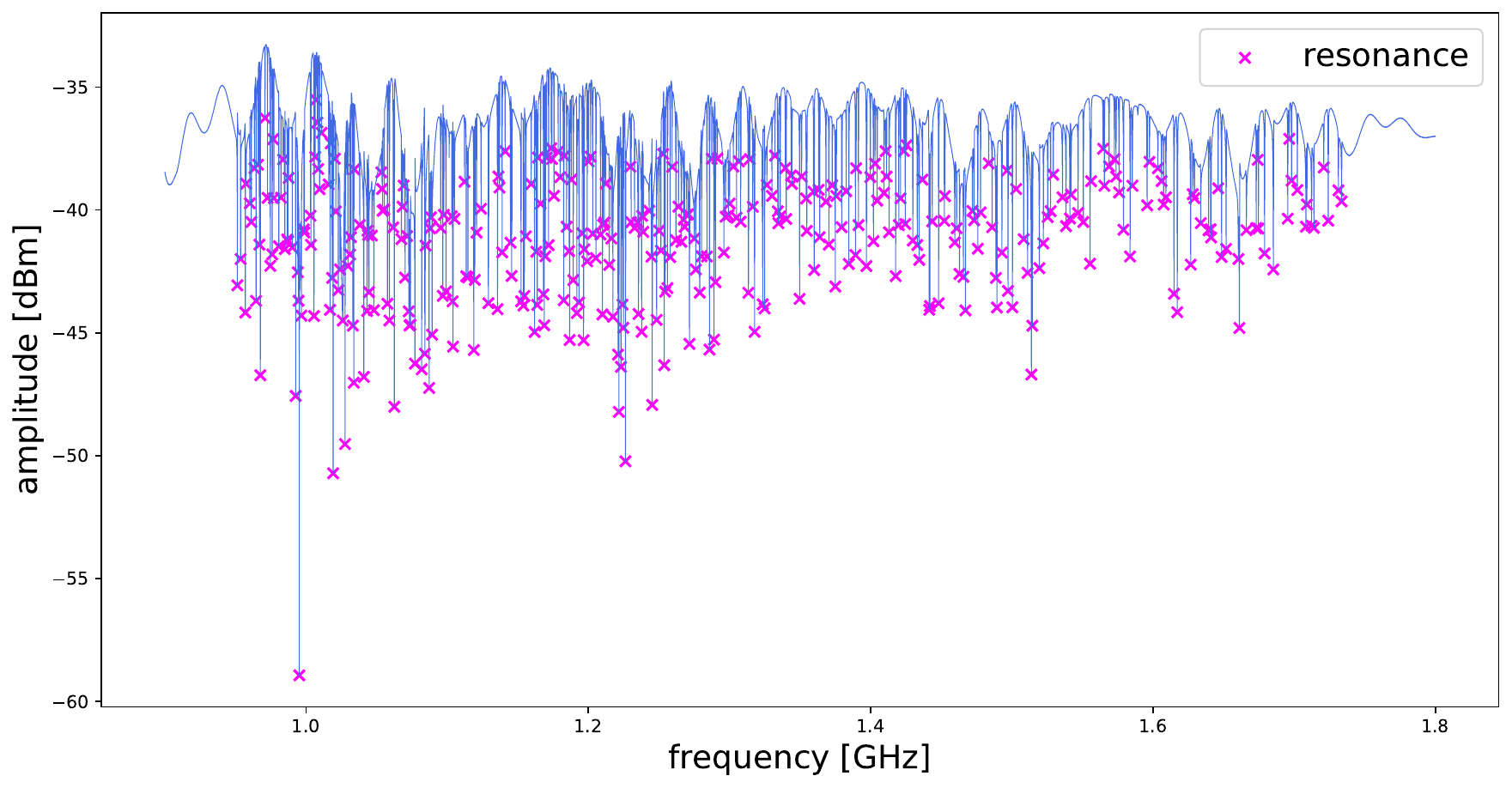}
    \end{subfigure}
    \caption{VNA feedlines of $AT$ (top) and $AR$ (bottom).}
    \label{fig:feedlines}
\end{figure}

To derive the main electrical characteristics of the arrays, we performed frequency sweeps using a VNA (Figs.~\ref{fig:feedlines}). The feedlines are not much affected by the presence of standing waves thanks to the bondings added across the feedline. 

This characterization has been focused on determining the internal, coupling and total quality factors of the arrays. The internal quality factor ($Q_i$) quantifies energy losses within the resonator, the coupling quality factor ($Q_c$) describes the energy transfer between the resonator and the external circuit, and the total quality factor ($Q_\text{tot}$) accounts for the combined effects of both. For the characterized arrays, resonances spanning 0.9 to 1.9~GHz exhibit $Q_{tot}$ values around  25 $\times 10^3$  under a representative optical background (see Tab.~\ref{tab:elec}). The measured quality factors align well with experimental predictions, confirming the design’s performance. Such predictions derive from previous experiments with similar optical load and pixel designs, such as NIKA and NIKA2 (\citealt{monfardini2011}, \citealt{monfardini2015}).

\begin{table}
\caption{Electrical characteristics of the 2~mm arrays $AT$ and $AR$.}
\centering 
\begin{tabular}{lcc} 
\hline\hline 
 & $AT$ & $AR$ \\ 
\hline 
Nº resonances & 398/418 & 345/418 \\ 
$Q_c$ & $(54 \pm 1) \times 10^3$ & $(56 \pm 6) \times 10^3$ \\
$Q_i$ & $(46 \pm 2) \times 10^3$ & $(49 \pm 4) \times 10^3$ \\
$Q_{tot}$ & $(24 \pm 1) \times 10^3$ & $(26 \pm 2) \times 10^3$ \\
\hline 
\end{tabular}
\label{tab:elec}
\end{table}

\subsection{Photometric and spectral characterization}
\label{sec:photo}

\begin{figure}
    \centering
    \includegraphics[width=\hsize]{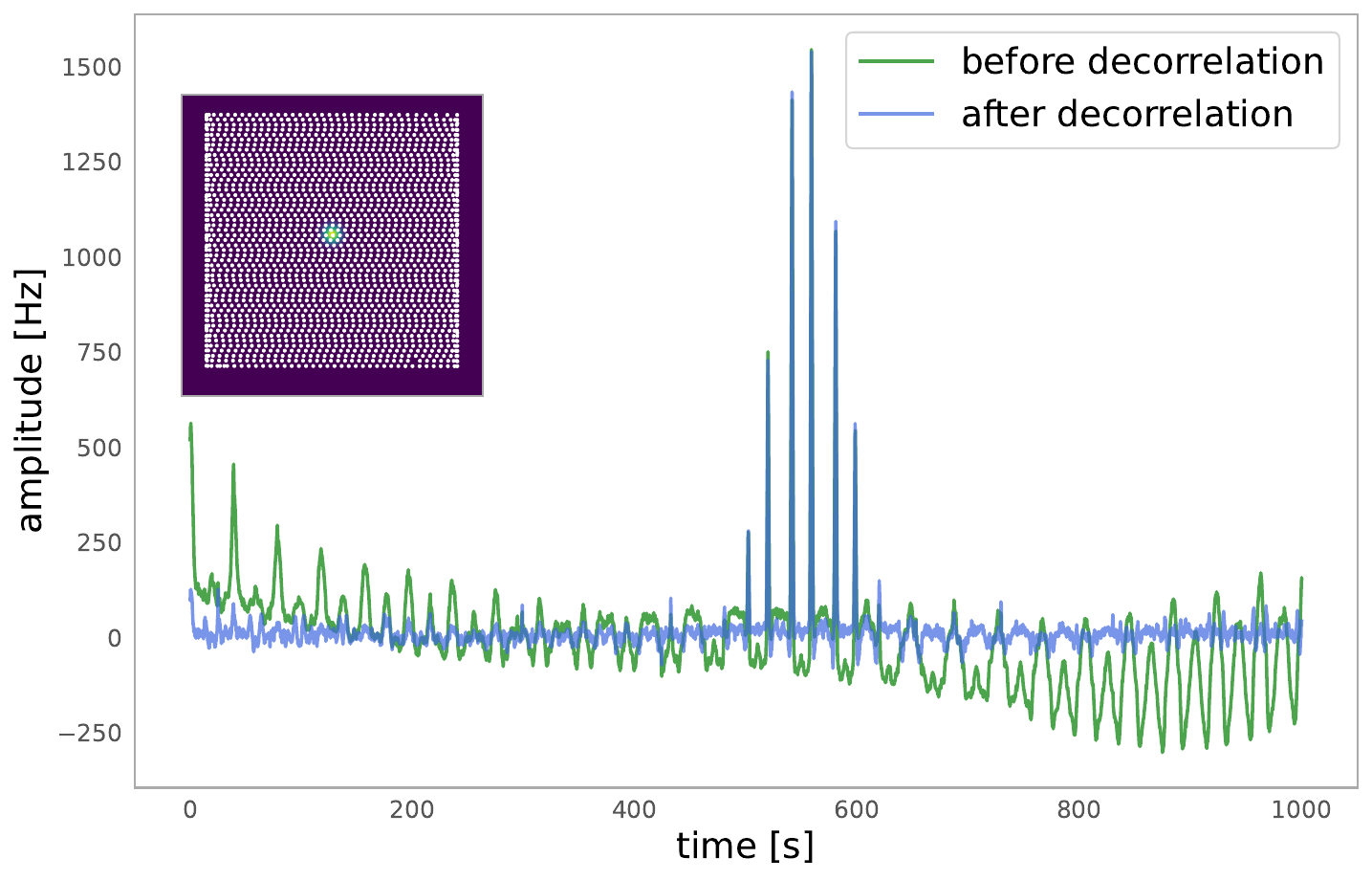}
    \caption{Typical timeline of the signal detected by one KID while the photometric source is being scanned, as shown in the top-left corner, in green before decorrelation and in blue after decorrelation. The intensity peaks correspond to successive passages of the photometric source through the line of sight of the selected KID, while the off-source oscillations in green result from small optical background variations following the scanning strategy.}
    \label{fig:ex_timeline}
\end{figure}

The data acquisition method involves performing on-the-fly (OTF) maps by raster scanning the photometric point source in front of the sky simulator, mimicking the mapping procedure used in telescopes. A raster scan covers a selected area with defined dimensions, subscan steps and scanning speed. Typically, the source is scanned over a range of $\pm$ 50~mm in both x and y directions, with a scanning speed of 4~mm/s and a subscan step size of 3~mm. This approach ensures uniform coverage of the mapped area while achieving diffraction-limited resolution for data analysis. With an expected beam FWHM of 14~mm and a subscan step size of 3~mm, the sampling factor is approximately 5, comfortably exceeding the Nyquist criterion of 2–3. A typical time stream of the signal detected by every KID is shown in Fig.~~\ref{fig:ex_timeline}, together with a typical raster scan in the top-left corner. In this plot, we distinguish the timeline before (green) and after (blue) subtracting a common mode of all KIDs and decorrelating, following the same method presented in \citealt{perotto2020}.

By investigating the bandwidth, Noise Equivalent Power (NEP) and beam profile, we aimed to validate the system’s design while assessing its sensitivity and efficiency under realistic operating conditions.

\begin{figure}
\centering
\includegraphics[width=\hsize]{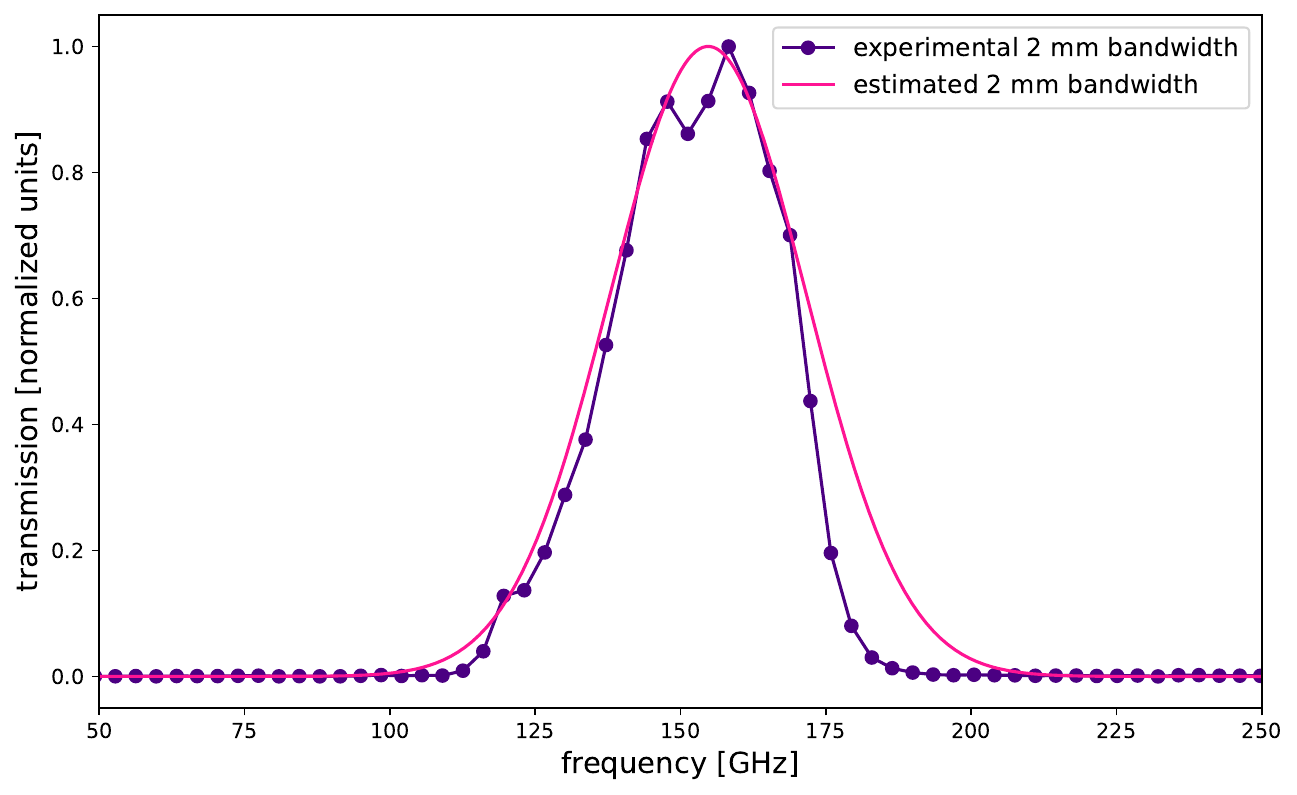}
  \caption{Normalized bandwidth for measurements at 2~mm: in blue the experimental bandwidth and in pink the estimated bandwidth, centered at 150~GHz and with a 30\% width as the experimental one.}
     \label{Fig:band_2mm}
\end{figure}

The bandpass, depicted in Fig.~\ref{Fig:band_2mm}, was determined from the absorption spectrum obtained using the Martin-Puplett interferometer configuration. For the optical estimations presented earlier (Sect.~\ref{sec:opt_sim}), the bandwidth was characterized by its fractional value, defined as $\Delta \nu / \nu_0$, where $\Delta \nu$ represents the width of the spectrum and $\nu_0$ is the central frequency. The measured bandwidth of approximately 30\% is consistent with typical photometric bandpasses, representing the 2~mm atmosphere window and confirming the suitability of the system for precision optical measurements.

To determine the single-pixel sensitivity averaged over the most responsive pixels in the array, we measured the optical responsivity. Two methods were employed to estimate the responsivity: one using the point-like photometric source and the other using an extended source created with an Eccosorb layer at 300~K. For the point-like source, the responsivity was determined as the ratio of the peak intensity value to the background level. In this case, the averaged responsivity over the usual selection of KIDs was measured to be $(0.25 \pm 0.02)~\mathrm{kHz/K}$ for the transmission array ($AT$) and $(0.24 \pm 0.02)~\mathrm{kHz/K}$ for the reflection array ($AR$). For the extended source, the responsivity was calculated as the resonance frequency shift divided by the temperature difference of the optical load, resulting in $(0.35 \pm 0.02)~\mathrm{kHz/K}$ for $AT$ and $(0.31 \pm 0.02)~\mathrm{kHz/K}$ for $AR$. In principle, the responsivity values obtained through the two methods should be identical if the optics were perfect. However, due to power losses in the far side lobes, the responsivity to the point-like source is reduced by approximately 30\%. For this study, we adopt the responsivity derived from the extended source, as our goal is to characterize the camera rather than account for optical losses.

Using the optical model illustrated in Sect.~\ref{sec:opt_sim}, the corresponding average power responsivity per pixel is found to be:

\begin{equation}
    \mathcal{R}_\mathrm{pixel} = (12.5\pm0.4)~\mathrm{kHz/pW}\, .
\end{equation}

\begin{figure}
\centering
\includegraphics[width=\hsize]{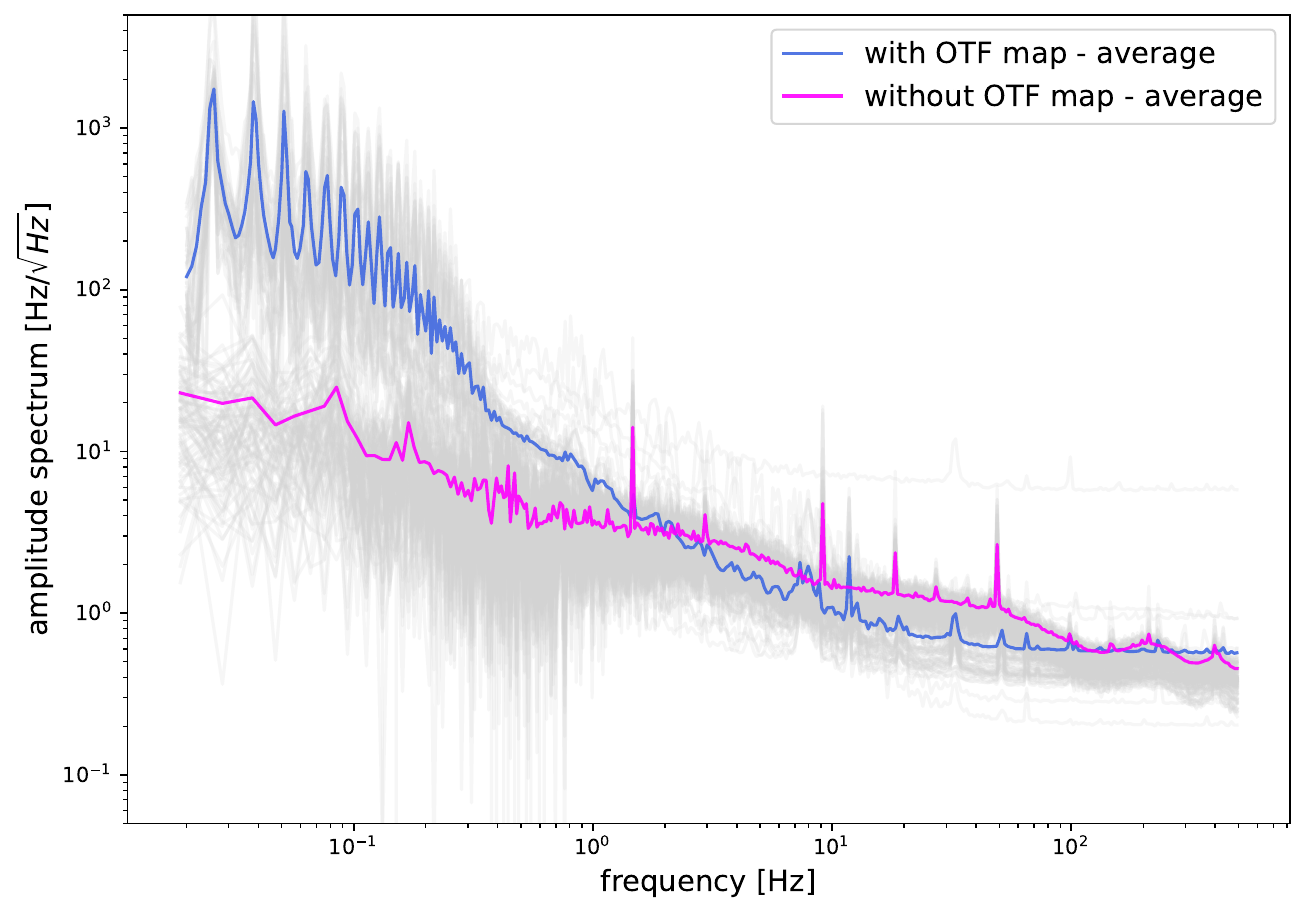}
  \caption{Comparison of spectra derived in two different conditions: in blue, we show the spectrum computed over data when the photometric source is being scanned, while in  magenta we show the pure instrumental noise. In gray the curves for each KID in the selection are shown.}
     \label{Fig:avg_pws}
\end{figure}

The spectral noise density $S_n(f)$ in Hz/$\sqrt{\text{Hz}}$ is calculated at a fixed sky simulator temperature of 22.3~K using a custom-made electronic system, with a multiplexing factor of 418 (\citealt{bourrion}). In Fig.~\ref{Fig:avg_pws} we compare the spectral noise densities obtained while an OTF map is running (in blue) and with the system stopped (in magenta), with no decorrelation applied to the timeline in neither case. The faint gray curves are the spectra of each KID of the 30\% selected KIDs, while the colored curves are the relative averaged values.

The $1/f$ noise detectable in the blue curve of Fig.~\ref{Fig:avg_pws} is caused by optical spatial fluctuations during a typical subscan, clearly visible in the green line of Fig.~\ref{fig:ex_timeline}. These fluctuations are due to the displacement of the optical source support, leading to micro-variations of the background temperature. The correlated optical noise is removed by subtracting a common mode before projecting the maps, again following the method of \citealt{perotto2020}. The noise spectrum acquired without running the OTF map does not exhibit a perfect $1/f$ behavior due to the presence of multiple fluctuation sources: optical, electronic and cryogenic. In particular, fluctuations in the optical background, essentially related to the temperature stability of the sky simulator, dominate the noise. These fluctuations are partially correlated across the KIDs array, as they observe similar behaviors; however, some noise components remain complex and difficult to disentangle. While the discrepancy in noise levels at low frequencies can be directly attributed to sub-scan movements and associated fluctuations, only present while the OTF map is running, the higher noise level observed without the OTF map running, compared to with it running, can likely be explained by differences in acquisition conditions. Specifically, the data were not collected simultaneously, and external noise sources may have varied between measurements. Nevertheless, the primary objective of this noise analysis was to confirm that the system operates within a realistic noise regime. Then, we selected the 1–10~Hz frequency range of OTF-free spectrum to compute the mean noise value, $<S_n(f)>$, equal to $(1.7\pm0.5)~\text{Hz}/\sqrt{\text{Hz}}$, because it corresponds to the typical timescale of the source passage through the beam, thus encompassing the relevant scientific signal. Although our setup currently does not include a rotating HWP, this frequency range remains appropriate when considering possible future project developments involving polarization modulation, as such modulation frequencies typically lie within this band.

Thanks to this noise analysis, we can compute the NEP for a pixel in the selected range 1-10~Hz as:

\begin{equation}
\text{NEP}_{\text{pixel}} = \frac{<S_n(f)>}{\mathcal{R}_{\text{pixel}}} = (1.5\pm0.5) \times 10^{-16} \, \text{W}/\sqrt{\text{Hz}}\,.
\end{equation}

This result satisfies Eq.~\ref{Eq:NEP_phot}: in fact, the relative discrepancy between $\text{NEP}_{\text{pixel}}(=\text{NEP}_{\text{goal}})$ and $\text{NEP}_{\text{phot}}$ is 1.1. This indicates that the difference between the values is comparable to their combined uncertainties, supporting the conclusion that we are actually dominated by the photon noise component.

Another result derived from the projected maps of the photometric source onto the focal plane is the estimation of the average beam. The beam is derived by projecting the time stream signal, as the one shown in Fig.~\ref{fig:ex_timeline}, onto a 2D map using the synchronous pointing data. 

\begin{figure}
    \centering
    \includegraphics[width=\hsize]{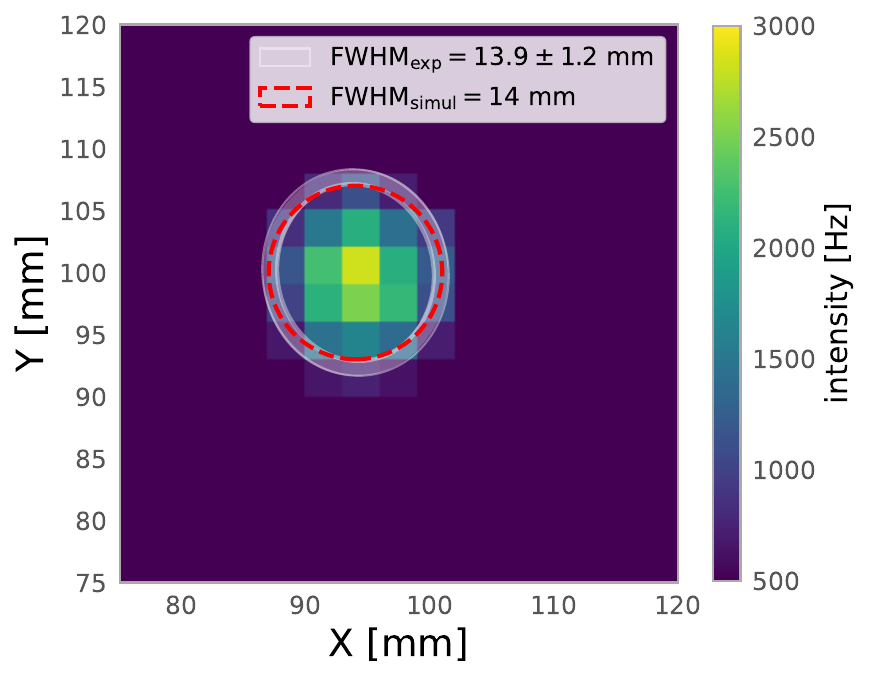}
    \caption{Projected image of the photometric source as seen by one sample KID. The white halo is the average FWHM over a selection of central KIDs of the focal plane (FWHM = 13.9 $\pm$ 1.2~mm, $e$ = 0.15 $\pm$ 0.01); the dashed red circle is the simulated FWHM = 14~mm, $e$ = 0.}
    \label{fig:beam}
\end{figure}

The distribution of beam sizes across the central region of the focal plane is uniform in size, with an average of FWHM = 13.9 $\pm$ 1.2~mm. By averaging the beams of a selection of central KIDs, we can see (in Fig.~\ref{fig:beam}) how closely the expected Zemax simulated FWHM approximates the experimental average FWHM, both in terms of mean value and ellipticity $e$.

\subsection{Polarimetry characterization}
\label{sec:polar}

Finally, we carried out polarization measurements using a 30~mm-diameter linear polarizer as the optical source positioned in front of the sky simulator. A precise reference for the source’s polarization angle was established empirically to serve as a baseline for comparison with the angle estimated by the detectors. The method involved aligning a polarizer’s wire with the wheel’s orientation, using a microscope to ensure precision given the wire’s small size (50~$\mu$m). By propagating the uncertainty over the entire length of the wheel, the initial orientation was determined with an accuracy of up to 0.1°.
This preliminary calibration ensures a precision of 0.1° for the orientation of the source polarizer (P1) relative to the optical line of sight.

Moreover, the polarizer at the 100~mK stage (P2) has been positioned with a given orientation of $\beta$ = 0$^\circ \pm 1$º, empirically leveled with respect to the line of sight. 

\begin{figure}
    \centering
    \includegraphics[width=\hsize]{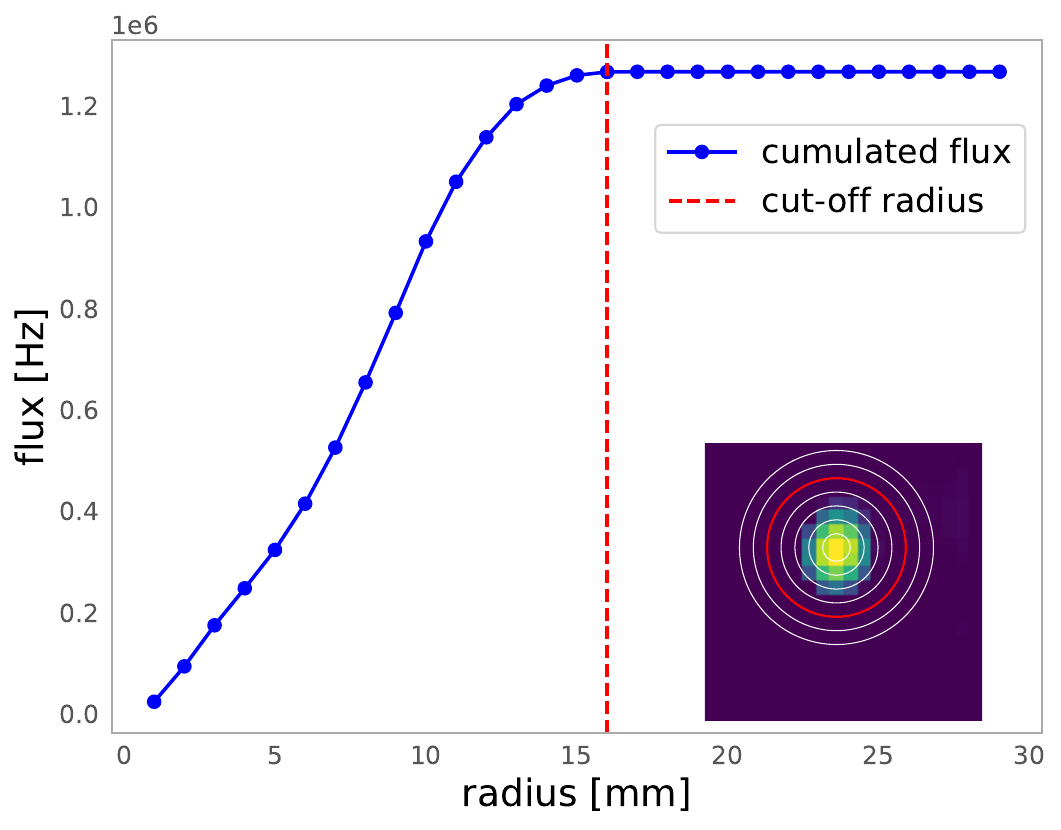}
    \caption{Aperture photometry technique used to estimate the cumulative flux within a defined radius. The red dashed line in the flux plot and the corresponding circle on the map represent the selected cut-off radius, indicating the boundary considered for analysis.}
    \label{fig:apert_photo}
\end{figure}

Polarimetric characterization involves the production of OTF maps in which for each map, the source polarizer (P1) is rotated by 15$^\circ$ relative to its previous orientation. This rotation is controlled via an angular encoder, enabling an angular precision of 0.01º (0.6~arcmin). By producing polarization maps for each KID analogous to those previously generated for a photometric point source, we applied an aperture photometry technique to extract the flux within a defined region of interest, as shown in Fig.~\ref{fig:apert_photo}. Aperture photometry is a widely used technique for estimating the flux of sources in maps, regardless of their specific morphology. This method requires defining two critical parameters: the center of the aperture and the cut-off radius. In this analysis, the center was selected as the location of the signal’s maximum intensity, ensuring the aperture is well-centered on the source of interest. The cut-off radius was determined to enclose 90\% of the total flux, providing a balance between capturing the majority of the signal and minimizing contamination from surrounding noise.

Using these flux measurements, we derived polarization curves averaged over the usual selection of best-performing pixels.  

\begin{figure*}
\resizebox{\hsize}{!}
        {\includegraphics[clip]{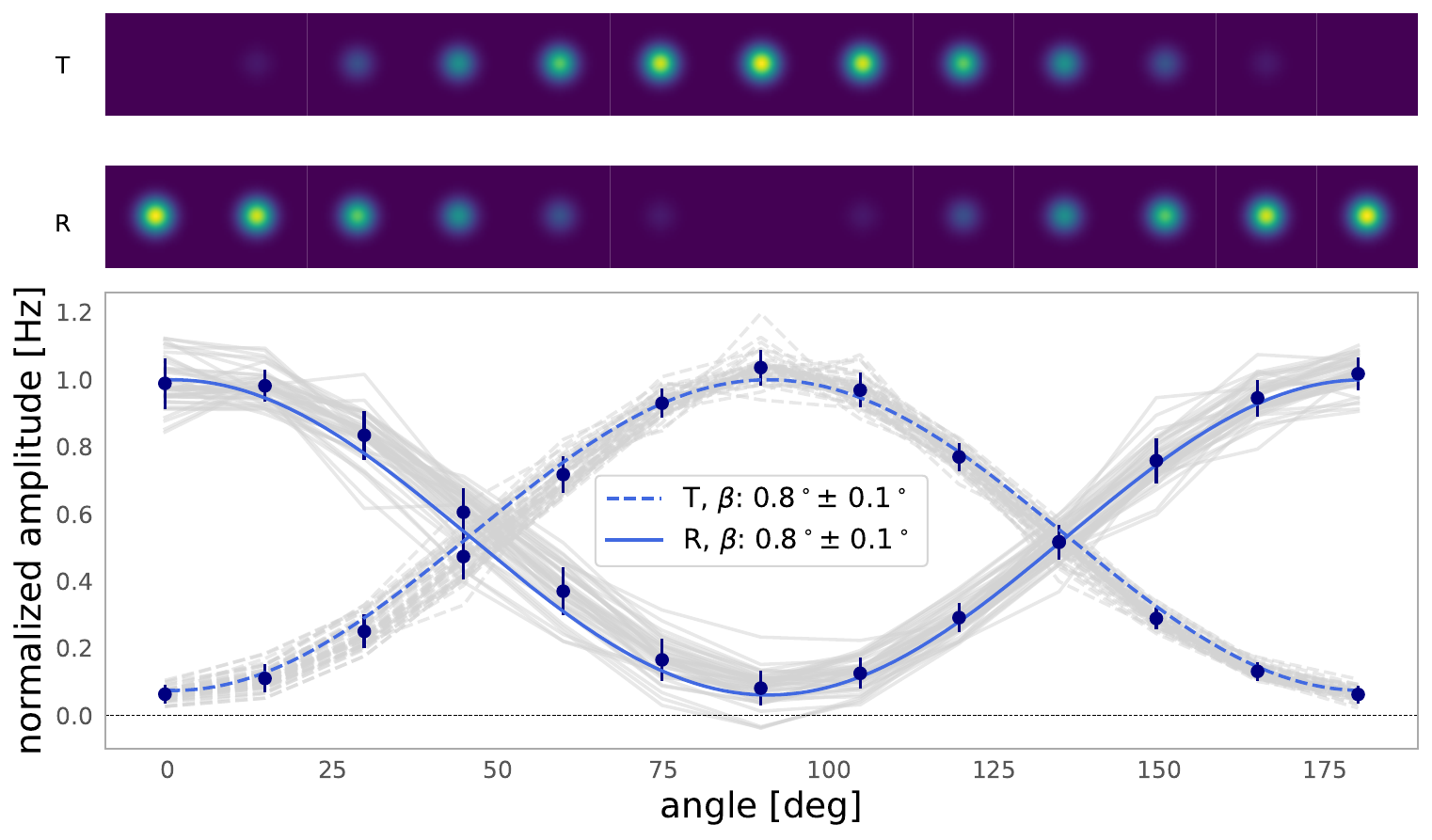}}
  \caption{Top panels: projected maps of the polarized source (P1) oriented at different angles as seen by the transmission (T) and reflection (R) arrays; Bottom panel: intensity of P1's maps as a function of the orientation angle for the transmission (T, dashed curves) and reflection (R, solid curves) arrays, obtained using an aperture photometry technique and fitted to derive $\beta = 0.8^\circ\pm 0.1^\circ$. The superposed gray lines represent individual results for the selection of best KIDs.}
     \label{Fig:polplot}
\end{figure*}

Fig.~\ref{Fig:polplot} shows the images of the polarized source (P1) in the focal plane as a function of its relative orientation with respect to P2. As expected, when the transmission array records a maximum signal, the reflection array observes a minimum, and vice versa. This behavior arises directly from the action of the cold linear polarizer, which splits the signal into two orthogonal polarizations, separated by a phase difference of 90$^\circ$.  

We can fit the measured data to an analytical model, in order to determine the orientation of P2 (angle $\beta$). The model function, $V$, shown in Eq.~\ref{eq:model}, was derived from the Stokes vector of unpolarized light, emitted by the sky simulator, and applying the Mueller matrices corresponding to the polarized source orientation (P1, angle $\alpha$) and the cold polarizer (P2, angle $\beta$): 

\begin{equation}
    V = 1 + \sin 2\alpha \cos 2\beta + \cos 2\alpha \sin 2\beta\, .
    \label{eq:model}
\end{equation}

The experimental data points in the bottom panel of Fig.~\ref{Fig:polplot} have been fitted jointly using a Monte Carlo Markov Chain (MCMC) algorithm implemented via the \texttt{emcee} Python library to ensure robust estimation of parameters (\citealt{emcee}). The fit results, obtained for the usual selection of best pixels, provide a final estimate of $\beta = 0.8^\circ\pm 0.1^\circ$. In the final error estimate for $\beta$, we have accounted for the intrinsic error on $\alpha$, 0.6~arcmin, the angular encoder’s uncertainty. This contribution is not shown in the plot as it is a minor and non-limiting source of error. The dispersion among the gray curves, each corresponding to a selected KID, arises from variations in their intrinsic response properties, differences in their positions on the focal plane, hence carrying optical distortions, and potential beam distortions. These factors may result in varying flux contributions when performing aperture photometry.

\section{Interpretation of results}
\label{sec:cosmo}

\subsection{Polarization angle accuracy}

The fitted estimate of $\beta$ can be interpreted in two ways, distinguishing between systematic and statistical uncertainties. 
The central value of $\beta = 0.8^\circ$, which is consistent with the expected value of $\beta = 0^\circ \pm1^\circ$, indicates that we have good control over systematic effects, especially considering the pre-calibration uncertainty on P1 and P2, which is limited to $1^\circ$. On the other hand, the associated statistical uncertainty of $\pm 0.1^\circ$ accounts for the reproducibility and the intrinsic dispersion observed among the KIDs within the same array, as represented by the superposed gray lines in Fig.~\ref{Fig:polplot}.

\subsection{Cross-polarization contribution}

\begin{figure}
    \centering
    \includegraphics[width=\hsize]{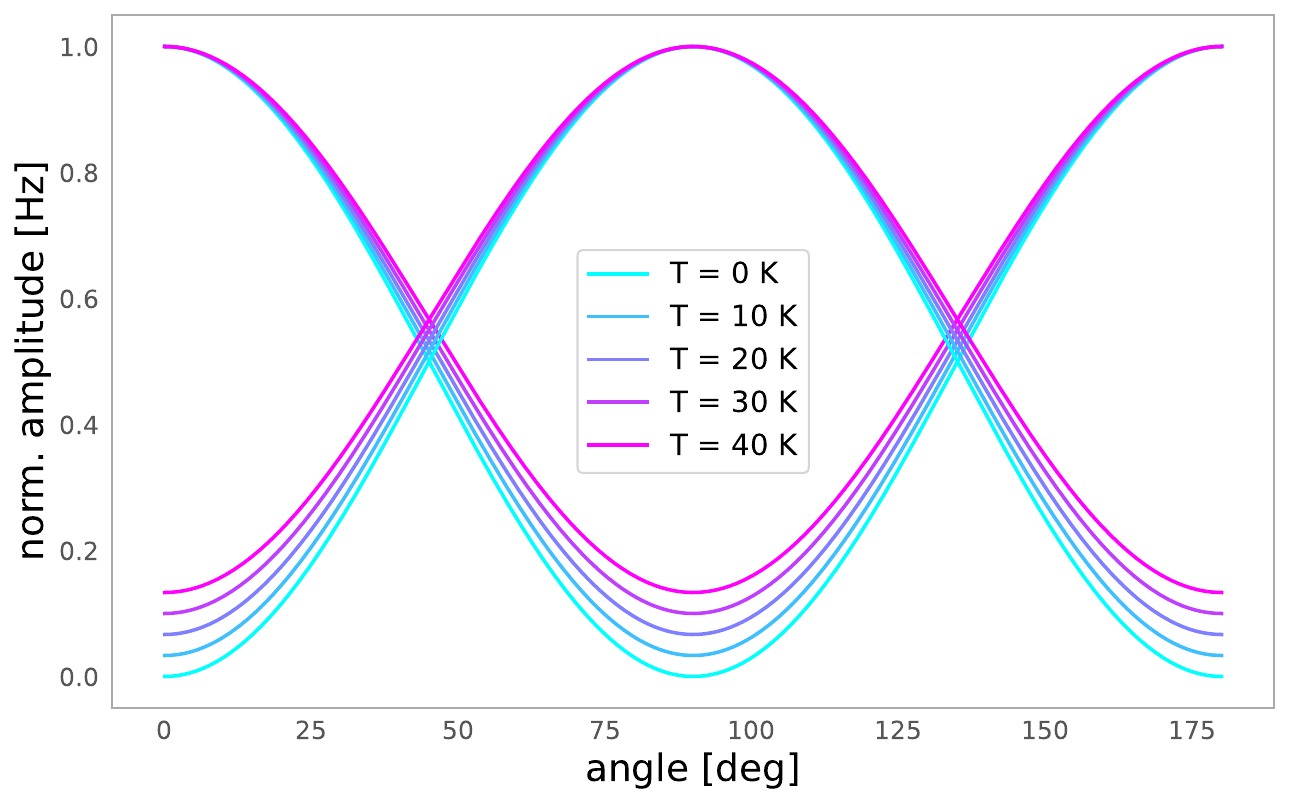}
    \caption{Simulated response of the curves as a function of the temperature of the reflected component, for both transmission and reflection arrays.}
    \label{fig:mod_double}
\end{figure}

Furthermore, it is important to note that the curves do not reach a zero level, which implies that the polarization is not entirely pure and there is an observable degree of cross-polarization. This effect results from observing two signal sources simultaneously: one from the sky simulator and another from the reflected ambient signal. The source polarizer (P1) does not emit any signal, but either reflects or transmits incoming signals based on its orientation. Positioned at a 10º downward tilt, P1 reflects the 300~K ambient signal while allowing the cold signal from the sky simulator to pass through. As a result, the 300~K signal dominates, but mixes with the cold sky simulator signal, preventing a complete nullification, even with the 90º phase difference between P1 and P2. This behavior is modeled by Eq.~\ref{eq:dual_model}:

\begin{equation}
    V_{dual} = V_{refl} + k(T) \cdot V_{trans}\, ,
    \label{eq:dual_model}
\end{equation}

where $V_{refl}$ is the reflection component, $V_{trans}$ is the transmission component, and $k(T) = T/300$ is a factor depending on the temperature. The model is represented in Fig.~\ref{fig:mod_double}.

Based on our analysis, we estimate that the parasitic component accounts for approximately 1\% of the signal, which corresponds to a cross-polarization of 1\% or parasitic temperature of around 17~K.

\subsection{Implications for CMB B-modes observations}
\label{sec:implic}

The tensor-to-scalar ratio $r$ quantifies the relative contributions of primordial gravitational waves, tensor modes, to the scalar perturbations observed in the CMB. Polarization measurements, particularly of the B-modes, are essential for constraining $r$. The uncertainty of the polarization angle introduces a systematic error in the measurement of the B-modes signal. This uncertainty can be represented as an angular error, $\Delta \psi$, in the reconstructed polarization angle. In this context, the uncertainty $\Delta \psi$ refers to the calibration error or uncertainty in the reconstruction of the polarization angle in the observed maps. This error can arise from inaccuracies in the instrument’s polarization angle calibration or systematic effects during data analysis. The effect of $\Delta \psi$ propagates into increased uncertainty in $r$, since the tensor-to-scalar ratio is directly influenced by the amplitude and angular precision of the observed polarization patterns. Quantifying the dependence of $r$ on $\Delta \psi$ allows us to assess the precision requirements for polarization angle measurements to robustly constrain $r$.

A miscalibration of the absolute polarization angle by $\Delta \psi$ introduces a coupling between the E- and B-modes of the polarization field. Specifically, the $C_\ell^{EE}$ and $C_\ell^{BB}$ power spectra, which describe the variance of the E- and B-mode polarization at a given angular scale $\ell$, become mixed. The power spectra \(C_\ell^{EE}\) and \(C_\ell^{BB}\) are defined as:  
\[
C_\ell^{EE} = \frac{1}{2\ell+1} \sum_{m=-\ell}^\ell \left|a_{\ell m}^E\right|^2, \quad 
C_\ell^{BB} = \frac{1}{2\ell+1} \sum_{m=-\ell}^\ell \left|a_{\ell m}^B\right|^2,
\]
where \(a_{\ell m}^E\) and \(a_{\ell m}^B\) are the spherical harmonic coefficients of the E- and B-modes, respectively. These coefficients are derived from the decomposition of the polarization field into curl-free (E-mode) and divergence-free (B-mode) components. The $C_\ell^{EE}$ power spectrum, typically much larger than the $C_\ell^{BB}$ power spectrum, leads to the phenomenon commonly referred to as "E-to-B leakage." The effect, as described in \citet{rosset}, can be quantified as shown in Eq.~\ref{Eq:cell}:

\begin{equation}
    \tilde{C}_\ell^{BB} = C_\ell^{BB} \cos^2(2\Delta \psi) + C_\ell^{EE} \sin^2(2\Delta \psi)\, ,
    \label{Eq:cell}
\end{equation}

where \(\tilde{C}_\ell^{BB}\) is the observed power spectrum in the B-modes, which includes a spurious component from the E-to-B leakage. The relative bias \(r\) is defined as:

\begin{equation}
    C_\ell^{BB} = r \cdot C_\ell^{BB}(r=1),
    \label{Eq:definition_r}
\end{equation}

where \(C_\ell^{BB}(r=1)\) represents the B-mode power spectrum in the absence of E-to-B leakage. Using this definition, \(r\) can be expressed as:

\begin{equation}
    r = \frac{\tilde{C}_\ell^{BB} - (2\Delta \psi)^2 C_\ell^{EE}}{C_\ell^{BB}(r=1)}\, ,
    \label{Eq:approx}
\end{equation}

which is defined over a given range of multipoles \(\ell\). For small miscalibration angles, the spurious contribution to the B-modes from the E-to-B leakage can be approximated as:

\begin{equation}
    \Delta C_\ell^{BB} \approx (2 \Delta \psi)^2 C_\ell^{EE}\, ,
\end{equation}

where \(\Delta C_\ell^{BB} = \tilde{C}_\ell^{BB} - C_\ell^{BB}\).

Eq.~\ref{Eq:approx} shows that the tensor-to-scalar ratio $r$ is sensitive to the polarization angle calibration error $\Delta \psi$. Larger values of $\Delta \psi$ lead to increased contamination from $C_\ell^{EE}$, which biases the estimate of $r$.

\begin{figure}
\centering
\includegraphics[width=\hsize]{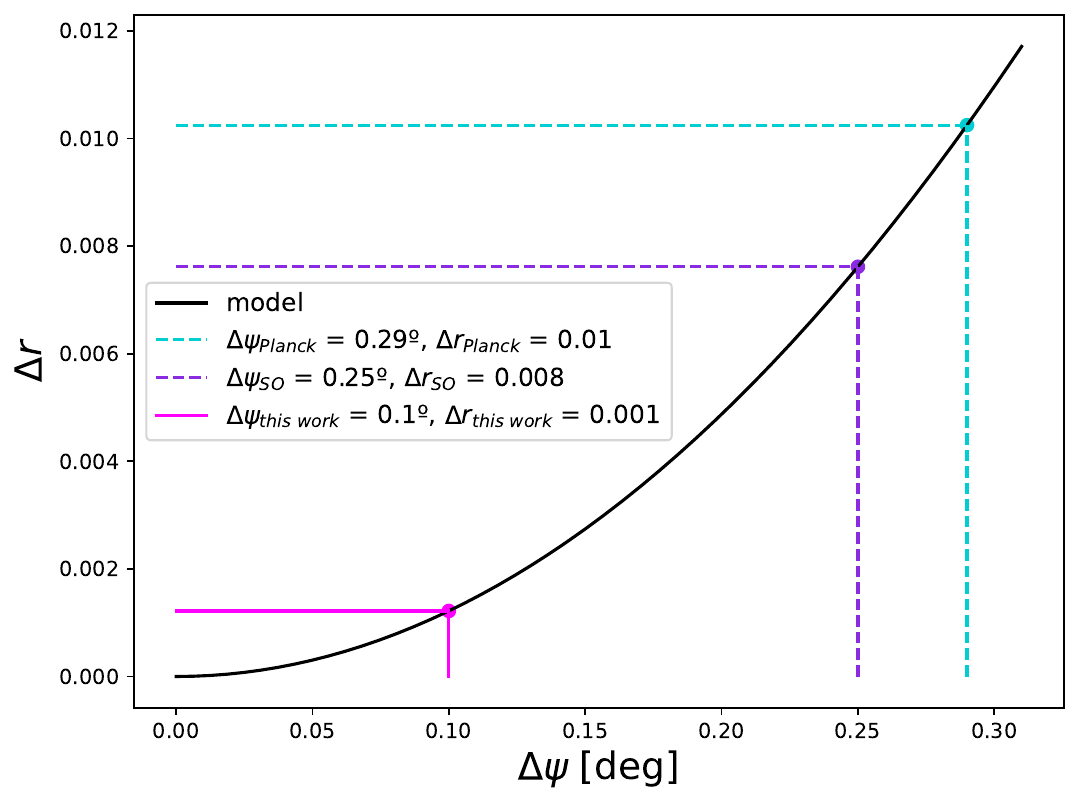}
  \caption{Forecast for $r$ values as a function of error on the polarization angle. The light blue line represents values from Planck constraint, the purple one for SO simulations and the magenta one shows results for this work.}
     \label{Fig:r_forecast}
\end{figure}

For a better understanding of the impact of $\Delta \psi$ on $r$, we can plot $\Delta r$ as a function of $\Delta \psi$, as shown in Fig.~\ref{Fig:r_forecast}. In this plot, we show the model curve, derived as explained in this section, compared to a set of experimental or simulated results. The range of $\ell$ taken into account is between 30 and 300, corresponding to the peak region of the $BB$ power spectrum. This estimate is coherent to previous results from Planck and predictions for SO (\citealt{aumont}). In the plot, we show upper limits from Planck intermediate results (\citealt{planck}), obtained considering the nulling effect from E to B modes. The forecast limit on $r$ for SO is also shown, following the simulated value presented in \citealt{giardiello}, combining estimates from the SO for both the Small Aperture Telescopes and the Large Aperture Telescope (\citealt{ade}). As we can see, our experimental uncertainty on the polarization angle lies well below all available and forecasted current constraints. 

\section{Conclusions}
\label{sec:concl}

This work has successfully established a fully equipped facility for rigorously testing the polarization capabilities of LEKIDs arrays. Using this setup, we demonstrated that two LEKIDs arrays, configured with a 45$^\circ$ linear polarizer between them, achieved polarization angle reconstruction with a precision of $0.1^\circ$, meeting the demanding requirements of next-generation CMB experiments. Additionally, the Noise Equivalent Power (NEP) of the arrays was measured to be $(1.5\pm0.5)\times10^{-16}$ W/$\sqrt{\text{Hz}}$, comparable to the photon noise limit ($\text{NEP}_{\text{phot}}$), highlighting the arrays’ competitiveness in millimeter-wave instrumentation.

The cryogenic system demonstrated stable operation with temperature fluctuations of the 100~mK stage within 1~mK during a typical measurement session. The optical components, including the cold polarizer, exhibited high reliability, while the detector arrays achieved the correct number of detected resonances and appropriate quality factors. The readout electronics ensured the required multiplexing factor, and the sky simulator demonstrated precision in angle and position encoding, with accuracy of 0.6~arcmin and 0.01~mm, respectively. The current uncertainty in polarization angle reconstruction is primarily influenced by the alignment and calibration accuracy of the experimental setup. This highlights the need for further refinements, particularly in improving the absolute reference characterization of the polarized source P1 and achieving precise alignment of the cold polarizer P2. These enhancements are expected to significantly reduce systematic uncertainties and improve measurement precision.

More generally, future efforts will include extending the validation of this concept to a larger selection of KIDs within the arrays. This will allow scaling up the number of detectors integrated into a single array for future experiments, aiming to achieve ideal uniformity across arrays containing up to 30k detectors. Additionally, we plan to expand the validation of these results from 150~GHz (2~mm) to higher frequencies, reaching up to 250~GHz (1.2~mm). A thorough investigation of beam distortions and their orientation-dependent effects is also critical to better understand their impact on the estimation of cosmological parameters. These advancements are vital steps toward qualifying this technology for its implementation in next-generation instruments utilizing LEKIDs technology. 

\begin{acknowledgements}
The authors acknowledge the contribution of LabEx FOCUS through the project ANR-11-LABX-0013 and of CNES through the R\&D project “PolarKID: mesures de polarisation du CMB par détecteurs LEKIDs”. This work was conducted in the context of the GIS KIDs in Grenoble (\url{https://gis-kids.cnrs.fr}).
\end{acknowledgements}

\bibliographystyle{aa}
\bibliography{aa54301-25}

\end{document}